\documentclass[pre,twocolumn,showpacs,preprintnumbers,amsmath,amssymb,superscriptaddress,floatfix,showkeys]{revtex4}

\usepackage{graphicx}
\usepackage{dcolumn}
\usepackage{bm}
\usepackage{color} 
\usepackage{amsfonts}
\usepackage{amsmath}
\usepackage{amsthm}

\begin{document} 

%\draft
 
\title{How P\'eclet number affects microstructure and transient cluster aggregation in sedimenting colloidal suspensions}

\author{A.\ Moncho-Jord\'{a}\footnote{ Electronic mail: moncho@ugr.es}} 
\affiliation{Biocolloid and Fluid Physics Research Group, Departamento de F\'{\i}sica Aplicada, Facultad de Ciencias, Universidad de Granada, Campus Fuentenueva S/N, 18071 Granada, Spain.}

\author{A. A. Louis\footnote{Electronic mail: ard.louis@physics.ox.ac.uk} }
\affiliation{Rudolf Peierls Centre for Theoretical Physics, 1 Keble Road, Oxford OX1 3NP, United Kingdom}

\author{J. T. Padding} 
\affiliation{Department of Chemical Engineering and Chemistry, Eindhoven University of Technology, P.O. Box 513, 5600 MB Eindhoven, The Netherlands}
 
\date{\today} 
 
\begin{abstract}
We study how  varying the P\'{e}clet number (Pe) affects the steady state sedimentation  of colloidal particles that interact through short-ranged attractions. By employing a hybrid molecular dynamics simulation method we demonstrate that the average sedimentation velocity changes from a non-monotonic dependence on packing fraction $\phi$ at low Pe numbers, to a monotonic decrease with $\phi$ at higher Pe numbers.  At low  Pe number the pair correlation functions are close to their equilibrium values, but as the Pe number increases, important deviations from equilibrium forms are observed. Although the attractive forces we employ are not strong enough to form permanent clusters, they do induce transient clusters whose behaviour is also affected by Pe number. In particular, clusters are more likely to fragment and less likely to aggregate at larger Pe numbers, and the probability of finding larger clusters decreases with increasing Pe number. Interestingly, the life-time of the clusters is more or less independent of Pe number in the range we study.  Instead, the change in cluster distribution occurs because larger clusters are less likely to form with increasing Pe number. These results illustrate some of the subtleties that occur in the crossover from equilibrium like to purely non-equilibrium behaviour as the balance between convective and thermal forces changes.
\end{abstract} 
 
\maketitle 

\section{Introduction}

The P\'eclet number (Pe) measures the ratio of convective to thermal forces.
 At low Pe numbers, where  thermal forces dominate, one might expect that properties such as the structure of a colloidal suspension would  be well approximated by their values at equilibrium (where Pe is strictly zero). On the other hand, at larger Pe numbers, purely non-equilibrium phenomena should become more prominent~\cite{Dhon96,Russ89}. 
 
 Nonetheless, the precise conditions under which the transition from equilibrium like to purely non-equilibrium behaviour occurs is a subtle question and depends on which properties are under investigation. For example, equilibrium correlation functions are often applied to calculate the properties of non-equilibrium systems. At what $\text{Pe}$ number does this approximation break down?

To address such questions, we study the interplay of thermal and convective forces in steady-state sedimentation, a classical non-equilibrium problem that, despite its apparent simplicity, it is still far from completely understood ~\cite{Dhon96,Russ89,Rama01}. A major difficulty for theories arises from the long-ranged ($1/r$) nature of the solvent induced hydrodynamic interactions (HI) that couple the motion of the particles in a complex way. Indeed, although Stokes \cite{Stok51} calculated the sedimentation velocity $v_s^0$ of a single hard sphere in 1851, it was necessary to wait more than 120 years until Batchelor \cite{Batc72} derived the first correction in the dilute limit, which is given by $v_s/v_s^0=1-6.55\phi$, where $v_s$ is the average sedimentation velocity of the colloids and $\phi$ is the particle volume fraction. Although this is strictly a non-equilibrium effect, it has been shown by simulations~\cite{Padd04} that for  hard sphere (HS) particles, the behaviour of $v_s(\phi)$ is virtually independent of Pe number down to at least $\text{Pe}=0.1$. So, for this example, a non-equilibrium effect persists virtually unchanged well into a regime where the thermal forces are significantly stronger than the convective forces.    
 
In this paper, we introduce a thermodynamic component to this non-equilibrium problem by studying the effect of short-ranged attractive interactions on colloidal sedimentation at different $\text{Pe}$ numbers. Nearly 30 years ago Batchelor calculated the effect of short-range interactions beyond the HS model on the average sedimentation velocity in the dilute limit \cite{Batc82a}, finding: 
\begin{equation}
\label{eq1}
v_s/v_s^0  = 1 - \left[6.55 - 3.52(1-B_2^*)\right]\phi + {\cal O}(\phi^2)
\end{equation}
where $B_2^* \equiv B_2/B_2^{HS}$,  $B_2$ is the second virial coefficient, and $B_2^{HS}$ is the virial coefficient calculated with the effective HS radius of the colloids. Eq.~(\ref{eq1}) suggests that adding attractions should increase the sedimentation velocity, while adding repulsions should decrease it compared to the pure HS case.  
An intuition for the increase in sedimentation velocity with attractions at low packing fractions can be obtained from the following considerations: Attractions should increase the probability of (transient) cluster formation. The sedimentation force scales linearly with the number  $i$ of particles in the cluster, whereas the friction is proportional to the radius of gyration of the cluster, which typically scales as $i^\alpha$, so that the sedimentation velocity of a cluster scales as $v_s \sim i^{1-\alpha}$. For compact clusters $\alpha \approx \frac13$ whereas for more open clusters $\alpha$ can be larger, although it cannot be greater than one. Even in the regime where the attractions are not strong enough to form permanent clusters, transient clusters can form, leading to a speed up in the average sedimentation velocity.

In the dilute limit, the increase in $v_s$ with increasing strength of attractions has been observed experimentally~\cite{Jans86,Plan09}. On the other hand, as the packing fraction increases, the gap between nearest neighbours decreases and so the fluid locally experiences an enhanced friction, which eventually slows down the average sedimentation velocity. What happens in between these two limiting regimes has been recently studied by Gilleland et al.~\cite{Gill11,Gill04}, who calculated $v_s(\phi)$ theoretically in the $\text{Pe} \ll 1$ limit, and who first predicted a maximum as a function of $\phi$ for strong enough attractions. We applied the same mesoscopic simulation technique we used for HS particles in~\cite{Padd04} to colloids with several different attractive interactions and also found that for strong enough attractions (but still below the point at which permanent clusters form) a similar maximum can be observed~\cite{Monc10}. These simulations were performed at just one intermediate $\text{Pe} = 2.5$, where the P\'eclet number is defined as $\text{Pe} = v_s^0a/D_0$, $a$ is the particle hydrodynamic radius, and $D_0$ is the equilibrium self-diffusion coefficient for a single colloid at infinite dilution. 

Here we investigate the effect of  varying the $\text{Pe}$ number at one fixed attractive interaction strength.
We find that the maximum in $v_s(\phi)$ v.s.\ $\phi$ is stable for $\text{Pe} \lesssim 1$, but starts to diminish at $\text{Pe}=2.5$ and has completely disappeared by $\text{Pe} = 10$. We also measure the colloid-colloid radial distribution function $g(r)$, and find that deviations from the equilibrium form begin to emerge for $\text{Pe} \gtrsim 1$.   

%Although permanent clusters don't form at equilibrium in our system, the attractions do promote transient cluster formation. We study the distribution of transient clusters as a function of Pe number and find that larger clusters are less common at larger Pe numbers. We also find that for a given cluster size, the number of fragmentation events grows with increasing Pe, while the number of aggregation events decreases with increasing Pe.

 The properties of the clusters formed in irreversible aggregation processes under sedimentation has been studied by means of Brownian Dynamics simulations in the region of low Pe number ($\text{Pe} <1$) in~\cite{Odri03,Odri04}. Although permanent clusters don't form at equilibrium in our system, the attractions do promote transient cluster formation.  Thus,  fragmentation combines with aggregation and sedimentation processes studied previously.  In contrast to the Brownian Dynamics simulations, we also include the effects of hydrodynamic interactions.  We find that increasing the Pe number reduces the number of larger clusters, and that for a given cluster size, the number of fragmentation events grows with increasing Pe, while the number of aggregation events decreases with increasing Pe.   

We proceed as follows: In the methods section we introduce our model potentials and describe the stochastic rotation dynamics computer simulation technique we employ. In the results section we then present and discuss the effect of Pe number on $v_s(\phi)$, $g(r)$ and on cluster distributions, and then summarize our findings in the final section.

\section{Model and Methods}

We employ a hybrid simulation method that treats the colloids with molecular dynamics (MD) and includes a background fluid that is by calculated with the Stochastic Rotation Dynamics (SRD) approach~\cite{Male99}. Such a hybrid technique was first employed by Malevanets and Kapral~\cite{Male00}, and we adapted it to study steady state sedimentation of HS particles in~\cite{Padd04}, and particles with attractive interactions in~\cite{Monc10}.
Moreover, the method has recently been shown to quantitatively describe colloidal sedimentation experiments, including complex nonlinear effects such as Rayleigh Taylor instabilities~\cite{Wyso09}. SRD itself (or Multi Particle Collision Dynamics as it is also called) has been fruitfully applied to a large number of other problems as well~\cite{Kapr08,Gomp09}. These previous successes give us confidence to apply this method to the problem of Pe number effects on sedimentation.

Within SRD, the fluid is not treated as a continuous solvent, but instead it is modelled by point-like particles. Thus the system is composed of $N_s$ solvent particles of mass $m_s$, and $N_c$ colloids of mass $M_c$. Collisions between solvent particles are efficiently coarse-grained in such a way that mass, momentum and energy are locally conserved, leading to the macroscopic Navier-Stokes equations~\cite{Male99,Kapr08,Gomp09,Padd06}. At the same time, Brownian motion of colloids is achieved naturally through colloid-solvent collisions. As SRD also includes thermal noise, the method represents a powerful technique to simulate driven colloidal suspensions embedded in a fluid. The choice of the simulation parameters must be done in a way that the most important dimensionless numbers fall in the correct hydrodynamic regime. Care must also be taken to ensure that physically relevant time-scales are properly represented. We refer to Refs.~\cite{Padd04,Padd06} for further information about the choice of the SRD parameters.

The interaction between solvent particles and colloids is represented by the repulsive WCA form~\cite{Hans86}
\begin{equation}
\label{HScf}
\beta V_{cs}(r)=\left\{\begin{array}{l@{\quad}l}
4\epsilon_{cs}\left[ \left( \frac{\sigma_{cs}}{r} \right)^{12}- \left( \frac{\sigma_{cs}}{r} \right)^{6} +\frac{1}{4} \right]  & r\leq 2^{1/6}\sigma_{cs} \\
0 & r > 2^{1/6}\sigma_{cs}
\end{array}  \right. 
\end{equation}
where $r$ is the distance between the solvent particle and the center of the colloid, $\beta=1/k_BT$, $\sigma_{cs}$ is the colloid-solvent interaction range and $\sigma_{cs}=2.5$. The use of radial colloid-solvent interaction implies that the fluid is not able to transfer angular momentum to the colloidal particle. Consequently, it induces slip boundary conditions at the colloid surface. Although it is also possible to generate algorithms with stick boundary conditions~\cite{Padd05,Whit10,Impe11}, we don't expect this to have any major qualitative affect on our results.

The colloid-colloid interaction has been modelled by a classic DLVO~\cite{Russ89} potential, where the total interaction is the sum of three contributions, $V_{cc}=V_{HS}+V_{vdW}+V_{DH}$. The first term is a repulsive hard sphere contribution given again by the WCA model but with a higher exponent to obtain a steeper hard-core repulsion
\begin{equation}
\label{HScc}
\beta V_{cc}(r)=\left\{\begin{array}{l@{\quad}l}
4\epsilon_{cc}\left[ \left( \frac{\sigma_{cc}}{r} \right)^{48}- \left( \frac{\sigma_{cc}}{r} \right)^{24} +\frac{1}{4} \right]  & r\leq 2^{1/24}\sigma_{cc} \\
0 & r > 2^{1/24}\sigma_{cc}
\end{array}  \right. 
\end{equation}
where $r$ now represents the distance between the centres of the colloids, $\epsilon_{cc}=2.5$ and $\sigma_{cc}=2.15\sigma_{cs}$ is the HS colloidal diameter. The colloid-colloid diameter was slightly larger than $2\sigma_{cs}$ to avoid spurious depletion forces between the colloids~\cite{Padd04,Padd06}. The second contribution is the repulsive Debye-H\"uckel interaction between the electrical double layers,
\begin{equation}
\label{DH}
\beta V_{DH}(r)=B\frac{\sigma_{cc}}{r}\exp[-\kappa(r-\sigma_{cc})]
\end{equation}
where $B=\frac{e^2}{4\pi k_BT\epsilon_0\epsilon_r\sigma_{cc}}\left[ \frac{Z_c}{1+\kappa \sigma_{cc}/2} \right]^2$, $Z_c$ is the effective charge of the colloid, $\epsilon_r$ and $\epsilon_0$ are the solution relative and vacuum dielectric constants respectively, and $\kappa$ is the reciprocal Debye screening length~\cite{Isra92}. Finally, the third term is the short-range London-van der Waals attraction~\cite{Russ89}, given by. 
\begin{equation}
\label{vdW}
\beta V_{vdW}(r)=-\frac{A_H}{12k_BT}\left[ \frac{\sigma_{cc}^2}{r^2}+\frac{\sigma_{cc}^2}{r^2-\sigma_{cc}^2}+2\ln \frac{r^2-\sigma_{cc}^2}{r^2} \right]
\end{equation}
where $A_H$ is the Hamaker constant. In order to overcome the singularity of the van der Waals contribution at contact, we also introduced a cut-off distance, $\delta$ (the so-called Stern layer)~\cite{Pell03}.

We study two different limiting situations: hard-sphere colloids and attractive colloids with a short-range attraction. In the former case, only the repulsive WCA potential was used. In the latter case, the potential parameters were given by $Z_c=8.3$, $\kappa=8.955/\sigma_{cc}$, $\beta A_H= 10.39$ and $\delta = 0.048\sigma_{cc}$. The temperature was assumed to be $T=293$K, and the relative dielectric constant of the solvent was given by the one for water, $\epsilon_r = 78.5$. The resulting reduced second virial coefficient was $B_2^* =-1.53$. To calculate $B_2^*$, we divided the virial coefficient through that of a hard sphere with an effective diameter equal to the range of the short-range repulsive barrier of $V_{cc}$. We note that there is some ambiguity in exactly how to obtain the effective diameter, as other recipes could be employed, but because of the steepness of the colloid-colloid interaction the differences are typically small, so we use this very simple one.

The equations of motion are updated with a standard Molecular Dynamics algorithm for the colloid-colloid and the colloid-fluid interactions, and with the coarse-grained SRD collision step for the fluid-fluid interactions. The number density of solvent particles was $\rho_s = 40/\sigma_{cs}^3$, the mass of the colloids $M_{c}=168m_s$, and the stochastic rotation angle was fixed to $\alpha = \pi/2$. This particular choice has been shown to reproduce correctly the hydrodynamic behaviour in the low Reynolds number regime, as well as the thermal Brownian fluctuations and diffusion for colloidal suspensions. The effective hydrodynamic radius obtained with these conditions is $a \approx 0.8 \sigma_{cs}$. We refer the reader to Refs.~\cite{Padd04,Padd06} for further discussion of the technical details of our SRD parameters.

The simulations were performed by placing  $N_c=8-819$ colloids inside an elongated box of sizes $L_x=L_y=16\sigma_{cs}$ and $L_z=48\sigma_{cs}$ with periodic boundary conditions in the three spatial directions. The number of solvent particles was $N=40V_{free}/\sigma_{cs}^3 \sim 3-5 \times 10^5$, where $V_{free}$ is the free volume left by the colloids. The sedimentation process is induced by applying a gravitational external field to the colloids in the $z$-direction. After an initial transient time, the system reaches steady-state conditions, where the average sedimentation velocity $v_s$ is constant, the colloidal microstructure does not depend on time, and no drift is observed in other measured observables either. In our simulations, we varied 
Pe from 0.5 to 10 by increasing the gravitational field. Averages were collected over runs from 5000 Stokes times $t_S$ (for Pe = 0.5) to $50000t_S$ (for Pe = 10), where $t_S = \sigma_{cs}/v_s^0$ is the time it takes an isolated sphere to settle one particle radius.  One could also define a Brownian time $\tau_B = \sigma_{cs}^2/D_0$, which is proportional to the time it takes the colloid to diffuse one radius.  Then $\tau_B \approx \tau_S \mathrm{Pe}$, and averages are taken over  periods ranging from $5000$ to $10000$ Brownian times, which should be ample for thermodynamic averaging.  We work in a frame where the downward volume flux of colloids is compensated by an equivalent upward volume flux of fluid.  In all cases the value of the Reynolds numberi is kept small  ($\text{Re}= v_s a/\nu < 0.4$, where $\nu$ is the kinematic viscosity), in order to ensure that the hydrodynamic behaviour of our system corresponds to the Stokesian regime typical for colloids. 

\section{Results for sedimentation velocities}

\begin{figure}
\center\resizebox{0.45\textwidth}{!}{\includegraphics{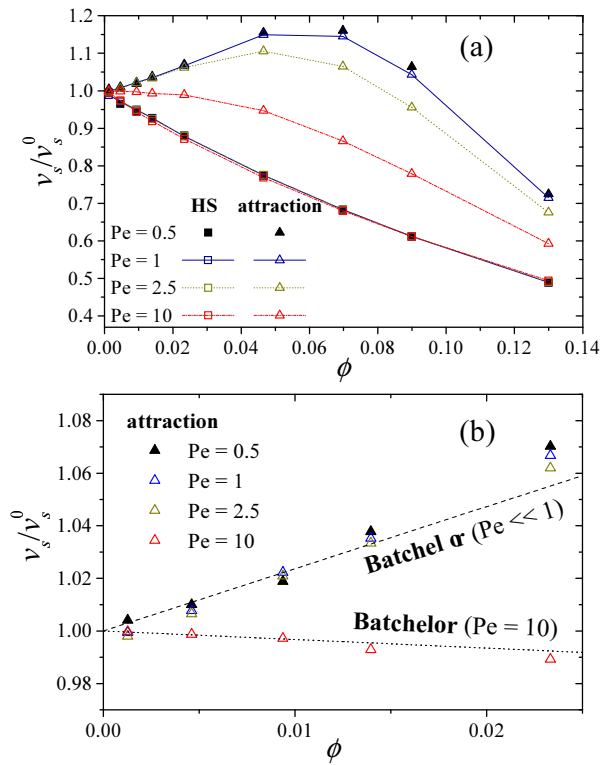}}
\caption{\label{Pe05_10} (a) Normalized average sedimentation velocity ($v_s/v_s^0$) as a function of the hydrodynamic packing fraction ($\phi$) for HS and attractive colloids, calculated for $\text{Pe}=0.5$, $1$, $2.5$ and $10$. (b) $v_s/v_s^0$ for attractive particles at the low density regime. The dashed line represents the Batchelor prediction given by Eq.~(\ref{eq1}) for the attractive system ($B_2^*=-1.53$) in the limit Pe$\rightarrow 0$. The dotted line applies Batchelor's low density limit equation, but employs the measured pair correlation function for Pe=10.} 
\end{figure}

In Fig.~\ref{Pe05_10}(a) we compare  $v_s(\phi)/v_s^0$  v.s.\  $\phi$  for the HS and attractive colloid systems at four different Pe numbers ($\text{Pe} = 0.5$, $1$, $2.5$ and $10$). The hydrodynamic packing fraction is given by $\phi=\frac{4}{3}\pi \rho_c a^3$, where $\rho_c$ is the colloidal number density. The hydrodynamic radius  $a$ is used because this sets the scale for the hydrodynamic interactions (one can also define a packing fraction based on colloid-colloid interactions, but this will typically be higher that the hydrodynamic one; the system is effectively non-additive~\cite{Loui01}). 
%Consider, for example, colloids interacting through long-ranged Coulombic repulsions.  Even at concentrations close to crystallization, there may still be substantial space between the particles for the fluid to pass through.  In other words, the hydrodynamic packing fraction will be significantly lower than the effective packing fraction derived from colloid-colloid interactions (the system is effectively non-additive~\cite{Loui01}).   

For hard spheres, as we increase the particle concentration, the hydrodynamic effect of the solvent leads to a rapid decrease of $v_s$, as first predicted by Batchelor~\cite{Batc72}. As found in~\cite{Padd04}, $v_s(\phi)/v_s^0$ is virtually independent of Pe number, and is in good agreement with the theoretical prediction of Hayakawa and Ichiki~\cite{Haya95} and with experimental data obtained for $\text{Pe} \rightarrow \infty$.    

The picture changes dramatically when we consider our system with attractive interactions.  On the one hand, for low Pe numbers, $v_s/v_s^0$ exhibits  non-monotonic behaviour, with a maximum at intermediate $\phi$, as predicted in~\cite{Gill04} and observed by us in previous simulations~\cite{Monc10}. According to Batchelor's pioneering calculations~\cite{Batc82a} for the dilute limit, shown in Eq.~(1), $v_s(\phi)$ increases with $\phi$ if $B_2^* \leq -0.86$, so our attractions are strong enough to induce an initial increase in $v_s(\phi)$ with $\phi$. Indeed, as shown in Fig.~\ref{Pe05_10}(b), at least for small Pe numbers ($\text{Pe} \lesssim 2.5$) the slope of the linear region at low $\phi$, i.e.\ the sedimentation coefficient, is  consistent with the Batchelor theory that, through Eq.~(\ref{eq1}), predicts that $v_s=v_s^0 \approx 1 + 2.364\phi$, which agrees well with our simulation data.

On the other hand, at high $\phi$, the fluid must pass through a dense collection of colloids and so we expect $v_s(\phi)$ to eventually decrease with increasing $\phi$.  The crossover between these two regions explains the maximum observed in Fig.~\ref{Pe05_10}(a).
 %This effect is the consequence of the competition between two phenomena, the hydrodynamics backflow retardation and the attraction-induced acceleration of the sedimentation \cite At low $\phi$ the formation of clusters induced by the attractions leads to an increase in $v_s$. Indeed, the gravitational force on a cluster grows linearly with the number of particles inside the cluster, but the friction increases roughly linearly with its radius of gyration. Since the latter typically increases less quickly than the former, the result is an acceleration of the sedimentation with the cluster size. 

For $\text{Pe} \lesssim 1$ the whole curve is independent of Pe, at least within the accuracy that we can measure. In contrast to what was found for HS systems, increasing the Pe number above this regime does change the behaviour significantly. For example, there is a progressive reduction of the maximum. For $\text{Pe} \ge 10$, this reduction is strong enough to completely remove the maximum. The dilute regime also becomes affected. For $\text{Pe} \gtrsim 2.5$ the sedimentation coefficient starts to noticeably decrease compared to the one predicted by the Batchelor.

\begin{figure}
\center\resizebox{0.45\textwidth}{!}{\includegraphics{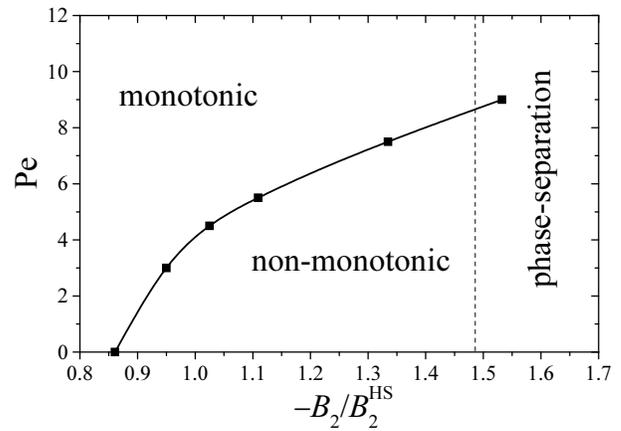}}
\caption{\label{Pe_B2} Different regions in the Pe--$B_2^*$ diagram: $v_s(\phi)/v_s^0$ either decreases monotonically with $\phi$, or else  shows a non-monotonic behavior (maximum). The diagram also has a third region for strong attraction where the colloidal suspension will phase separate and exhibit permanent clusters.  $B_2^* \lesssim -1.5$ is a good lower bound for where this behaviour will begin to set in~\protect\cite{Vlie00}.} 
\end{figure}

The critical value of the P\'eclet number that leads to this non-monotonic behaviour depends on the interaction potential, but for short-ranged interactions a maximum can occur for sufficiently small Pe numbers if $B_2^*<-0.86$. As Pe increases, we expect the maximum to disappear. Preliminary simulations at other $B_2^*$s suggest that the weaker the attractions, the lower the critical Pe number at which the monotonic behaviour sets in.  On the other hand, for strong enough attractions, permanent clusters will form and lead to different physics than what we are considering here. It is known that the criterion $B_2^* \lesssim 1.5$ provides a remarkably accurate prediction of the potential strength at which a critical point sets in~\cite{Vlie00}. Of course the critical point is only at one concentration, so one could potentially reach a slightly more negative $B_2^*$ at lower concentrations as long as the fluid-fluid binodal or spinodal towards phase separation (and permanent cluster formation) is not crossed (note that the location of these lines will be affected by the sedimentation as well). One must also take care not to cross a fluid-solid binodal or other phase-lines.  We illustrate the region of non-monotonic behaviour with a schematic diagram in Fig.~\ref{Pe_B2}.

%i find a critical Pe that separates these two regions.
 %For short-range attractions the second virial coefficient gives a fair estimate of the interparticle attraction strength. 
 %Fig.~\ref{Pe_B2} shows the different regions in the $\text{Pe}-B_2^*$ diagram where $v_s(\phi)/v_s^0$ decreases monotonically, or shows a maximum. For $B_2^*>-0.86$, $v_s(\phi)$ is always a decreasing function of $\phi$, but for $B_2^*<-0.86$ a maximum is expected to occur at small Pe numbers.   However, upon increasing 
 
  %it is always possible to find a critical Pe that separates these two regions. For small attractions, the critical Pe is close to zero, and grows with the attraction strength. For $B_2^*<-1.5$ the short-range attractive bonds between colloids becomes strong enough to induce the formation of permanent clusters~\cite{Vlieg00} and the systems  can phase separate.

\section{Results for microstructure}

\begin{figure}
\center\resizebox{0.45\textwidth}{!}{\includegraphics{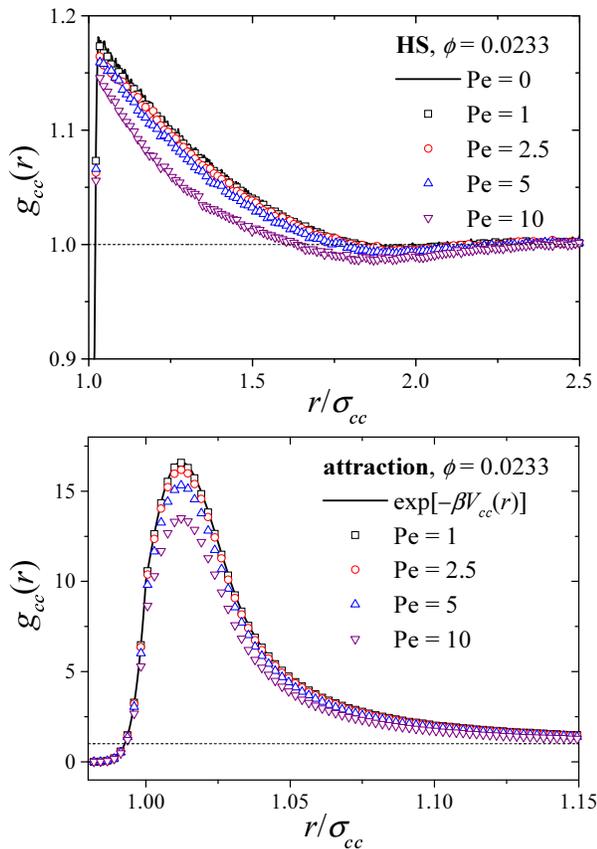}}
\caption{\label{gr} Radial distribution function $g(r)$ as at different Pe number for HS (top) and attractive colloids (bottom). Note the differences in scale between the top and bottom graphs.} 
\end{figure}

In order to investigate further the possible causes for the change of $v_s(\phi)/v_s$ with Pe, we also studied the microstructure of the suspensions, including radial distribution functions and cluster distributions.

%At this point, it is important to emphasize that these new P\'eclet number effects are necessarily linked to changes in the colloidal microstructure. At low Pe, the convective flow induced by the gravity field is too weak to substantially alter the structural properties of the colloidal system (radial distribution function, cluster distribution,...), so that they can be approximated by the equilibrium ones. This explains the saturation of $v_s/v_s^0$ for $\text{Pe} \leq 1$. At high Pe, the flow around the particles and clusters affect the spatial distribution of colloids compared to the one reached at equilibrium conditions and so, modify the average sedimentation rate.

Fig.~\ref{gr} depicts the radial distribution function of the settling colloids at five different Pe numbers (from $\text{Pe}=1$ to $10$) for HS and attractive colloids at $\phi = 0.0233$.  For $\text{Pe}  \lesssim 1$ the radial distribution functions for both HS and attractive colloids do not vary appreciably with Pe, and agree with the equilibrium distributions (at $\text{Pe} = 0$).  For larger Pe numbers we observe a small deviation for the HS correlation functions, including the emergence of a small but distinct hydrodynamically induced anti-correlation region around $r \approx 1.8\sigma_{cc}$ for $\text{Pe} = 10$.  Note that the scale of the changes is very small, so that the effect on $v_s(\phi)$ is expected to be very modest as well, most likely well within the error bars of our calculation of sedimentation coefficients we measure.  

\begin{figure}
\center\resizebox{0.45\textwidth}{!}{\includegraphics{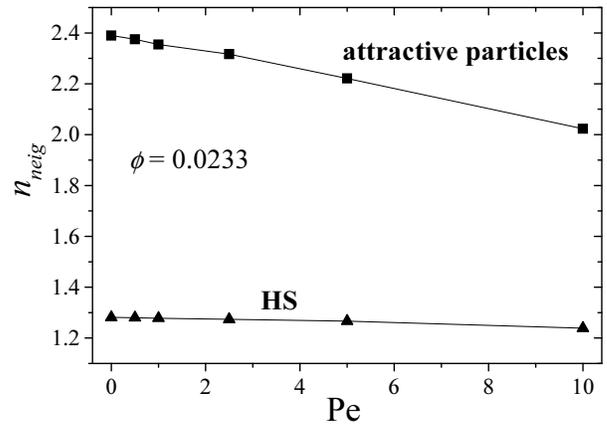}}
\caption{\label{neig} Average number of neighbours $n_{neigh}$, given by Eq.(\protect\ref{eq_neigh}), 
around a colloid as a function of Pe for HS and attractive particles.} 
\end{figure}
 
In contrast to the case for HS particles,  for attractive interactions the effect of Pe number on the pair correlations is much more pronounced.  For Pe $\lesssim 1$, $g_{cc}(r)$ is indistinguishable from its $\text{Pe}=0$ form, which at these low packing fractions can be accurately reproduced by the dilute limit expressions $\exp {[-\beta V_{cc}(r)]}$~\cite{Loui01a}. As we increase the Pe number above 1, the main peak of $g_{cc}(r)$ begins to noticeably decrease in magnitude.  At Pe=10, the deviations are quite substantial.  One measure of this hydrodynamic effect can be quantified by measuring the average number of nearest neighbours around a particle, which is given by the integral of $g(r)$ from $r=0$ to the position of the first minimum(located at $r \approx 1.5\sigma_{cc}$):
\begin{equation}
\label{eq_neigh}
n_{neig}=4\pi\rho_c \int_0^{1.5\sigma_{cc}}g_{cc}(r)r^2dr.
\end{equation}
As shown in Fig.~\ref{neig},  $n_{neig}$ drops from 2.39 to 2.02 as we increase Pe from 1 to 10. In other words, here the effect of increasing the relative strength of the convective over thermodynamic forces reduces the probability that two particles are in close proximity.  

%For HS, it is interesting to note that the structure shows an anticorrelation region around $r \approx 1.8\sigma_{cc}$ for $\text{Pe} = 10$. This is a pure hydrodynamic effect that indicat es the existence of a lubrication layer around the colloids acting as an effective repulsive interaction. Nevertheless, this effect is rather small. Indeed, by calculating $n_{neig}$ using the same integration limits, we find that it goes from 1.28 to 1.24 as we increase Pe from 1 to 10, what represents only a change of about $3\%$.

%Since the Batchelor theory uses the equilibrium distribution function to calculate the sedimentation rate, these results explains why Eq.~(\ref{eq1}) overestimates the sedimentation coefficient for attractive colloids at low $\phi$. This example illustrates the danger of using equilibrium correlation functions in non-equilibrium calculations.

In the $\rho \rightarrow 0$ limit, instead of the equilibrium correlation function $g(r) \approx \exp (-\beta V(r))$, one could also use the measured non-equilibrium $g(r)$ in the derivation leading to Eq.~(\ref{eq1}). When we use the measured $g(r)$ for Pe = 10, we find that $v_s/v_s^0 \approx1-0.326\phi$. The sedimentation coefficient now has the opposite sign to that found when using the $g(r)$ at Pe = 0, and in fact fits the data for P = 10 reasonably well, as can be seen in Fig.~\ref{Pe05_10}(b). This example illustrates the danger of using equilibrium correlation functions in non-equilibrium calculations.

%Since the Batchelor theory uses the equilibrium distribution function to calculate the sedimentation rate, these results explain why Eq.~(\ref{eq1}) overestimates the sedimentation coefficient for attractive colloids at low $\phi$. Here we observe an good illustration of the potential pitfalls of employing equilibrium correlation functions in non-equilibrium calculations. Nevertheless, it is still possible to make use of the Batchelor model at large Pe by replacing the equilibrium distribution by the real non-equilibrium one. For instance, one can calculate an effective pair potential for Pe=10 from the knowledge of the simulated pair distribution function in the dilute regime by simply taking $\beta V_{cc}(r)=-\ln(g_{cc}(r))$. The resulting effective second virial coefficient is given by $B_{2,eff}^* = -0.62$, which is above $-0.86$, as expected. Using Eq.~(\ref{eq1}) we can obtain a prediction for the dilute regime, $v_s/v_s^0=1-0.85\phi$. This is plotted in Fig.~\ref{Pe05_10}(b) as dotted lines. As observed, the agreement with the simulation results for Pe=10 is fairly good, indicating that the Batchelor model can be extended to non-equilibrium situations by considering the real pair distribution function.  

Next we focus on another measure of microstructure, namely cluster distributions.  Two particles are considered to be in a cluster if they are within a cutoff radius of $r=1.06 \sigma_{cc}$ of each other, which places them well inside the attractive potential well. Other cutoffs could be used, but these don't qualitatively change the behaviour we observe. The clusters are transient, that is they form and break up due to thermal and convective forces. Moreover, in steady-state, the average cluster distributions should not change with time (although fluctuations may occur caused by finite size effects). Using this criterion, we measured the probability distribution of transient clusters $P(i)$ and their average life-time $\tau_i$ as a function of the cluster size, $i$. We used the same criterion to calculate cluster distributions for HS particles.
 
\begin{figure}
\center\resizebox{0.45\textwidth}{!}{\includegraphics{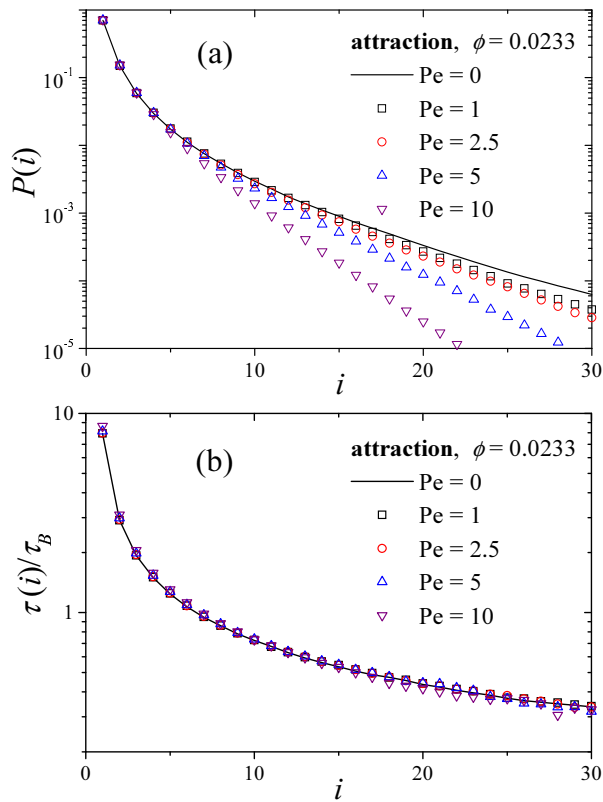}}
\caption{\label{CSD_Pe} The plots show (a) the probability of finding a cluster of size $i$ and (b) the average cluster life-time (normalized by the Brownian time $\tau_B$) for attractive particles at several Pe numbers. In all cases the packing fraction of colloids is $\phi=0.0233$.}
\end{figure}

\begin{figure}
\center\resizebox{0.45\textwidth}{!}{\includegraphics{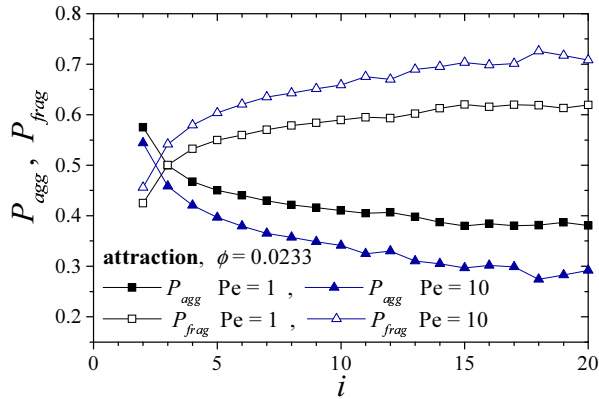}}
\caption{\label{Pagg} Probability of aggregation and fragmentation events calculated for attractive particles at two P\'eclet numbers, $\text{Pe}=1$ and $10$. The packing fraction of colloids is again $\phi=0.0233$.}
\end{figure}

For HS particles, both properties are only weakly affected by Pe number over the range we study. In contrast, for attractive particles, Fig.~\ref{CSD_Pe}(a) demonstrates that increasing Pe leads to a substantial decrease of the probability of finding large clusters. At low packing fractions, this may help explain the decrease of the average sedimentation coefficient with increasing Pe number, as larger clusters are expected to sediment faster. 
%The reason for the changes in the cluster distribution is that the larger convective forces break up the clusters more easily. In fact, we do not really need to reach extremely large values of Pe to observe the effect of the flow on the transient clusters since the interparticle attractions are not very strong. For more attractive systems (where permanent clusters are formed) it will be necessary to increase much more Pe in order to induce the cluster break up.

Given that there are fewer large clusters with increasing Pe number,  and that convective forces increase with Pe, one might  also expect a decrease of the the cluster-life times as Pe grows. However, as can be seen in Fig.~\ref{CSD_Pe}(b), for $\phi = 0.0233$, the cluster life-times are nearly independent of Pe number over the range of Pe numbers we investigate. We note that this surprising effect seems limited to very low $\phi$. At $\phi \approx 0.05$, we do observe a decrease in cluster life-times with increasing Pe. To explain why the average life-times of the clusters does not change at low $\phi$ it is important to think further about the processes that lead to a certain cluster size. These include both aggregation processes, whereby smaller clusters coagulate into larger ones, and fragmentation processes where a larger cluster breaks up into smaller ones. The larger the cluster, the more different formation and breakup pathways that are available. The life-time of a cluster can decrease both because the rate of aggregation increases or because the rate of fragmentation increases.  

To study the statistics of fragmentation and aggregation, we tracked, over a suitably long averaging time, the number of times a given cluster of size $i$ decreased in size due to fragmentation events, $n_{frag}(i)$ as well as the number of times it increased in size due to aggregation events, $n_{agg}(i)$. These numbers increase linearly with time, so it is useful to define normalized  quantities $P_{agg}(i)=n_{agg}(i)/n(i)$ and $P_{frag}(i)=n_{frag}(i)/n(i)$, where $n(i)$ is the total number of times that a cluster of size $i$ emerges during the averaging time.  In steady state, $P_{agg}(i) + P_{frag}(i) = 1$, so that each can be viewed as a probability that a cluster of size $i$ disappears by either fragmentation or aggregation respectively. These probabilities are plotted in Fig.~\ref{Pagg} for Pe=1 and Pe=10. We see firstly that the probability of aggregation decreases with increasing cluster size, whereas the rate of fragmentation increases with increasing $i$, as one might expect because there are more ways a large cluster can break up. Moreover, these probabilities change with Pe number. For higher Pe number the probability of fragmentation increases, as one might expect due to the increased relative strength of shear forces over thermal forces.  However, the probability of aggregation also decreases due to the flow.  The combination of these two rates apparently balance each other out, and so the average life-time of a cluster at $\phi = 0.0233$ appears not to change with increasing Pe number. At higher packing fractions these cluster life-times do change. By contrast, the increase in the probability of fragmentation, combined with the decrease in the probability of aggregation does lead to a decrease in the {\em number} of larger clusters. Clearly the effect of changing the balance of convective to thermal forces on the aggregation and fragmentation of clusters is complex. A fuller investigation would not only track the cluster size distributions, but also the different pathways of formation/breakup. A future publication will investigate these effects in more detail.

\section{Discussion}

In summary, we have used a mesoscopic simulation technique to study the effect of changing the balance between thermal and convective forces, quantified by the Pe number, in a sedimenting suspension of colloidal particles.  Whereas for pure HS particles the effects of varying the Pe number between 0.1 and 10 are quite small ~\cite{Padd04}, for attractive particles there is a clear change in behaviour over this range of Pe numbers.  Firstly, the maximum in the sedimentation velocity $v_s(\phi)$ with $\phi$, first predicted in~\cite{Gill04,Gill11}, and observed by simulations in~\cite{Monc10}, disappears as the Pe number increases.  Secondly, the radial distribution function $g(r)$ changes noticeably for Pe $\gtrsim 1$:  increasing the Pe number means the average number of nearest neighbours decreases compared do the equilibrium value. Thirdly, the cluster distributions change: the probability of observing larger clusters decreases with increasing Pe.  Moreover, increasing the Pe number increases the probability that a transient cluster fragments, and decreases the probability that it aggregates with another particle or with other clusters. This work illustrates the often complex crossover from equilibrium-like to purely non-equilibrium behaviour as the balance between thermal and convective forces changes.

A number of further questions are raised by this study.  First of all, we only used one fairly short-ranged potential form. It would be interesting to see how these results depend on the shape of the potential. For example, a longer-ranged attractive potential with the same integrated strength, which through Eq.~(\ref{eq1}) would generate a similar sedimentation coefficient, would have a less deep dimensionless well depth $V_{min}/ k_B T$. One might expect that finite Pe number effects set in roughly when  $V_{min}/k_B T \sim$ Pe, i.e. when the shear forces are strong enough to dislodge particles that cluster together \cite{Boek08}. Thus a for a longer-ranged potential, finite Pe number effects should set in earlier than for a shorter-ranged potential. Similarly, it would be interesting to investigate what the effect of Pe number is on longer-ranged repulsive potentials~\cite{Thie95}.   
  
Another interesting direction of investigation would be to study what happens at larger interaction strengths where permanent clusters can form.  The shape of clusters~\cite{Whit11} will most likely also depend on Pe number, and we anticipate a rich physics as a function of the interplay between the aggregation of clusters and the convective forces driven by their sedimentation.

\acknowledgements

The authors thank the Spanish Ministerio de Educaci\'on y Ciencia (project MAT2009-13155-C04-02), the Junta de Andaluc\'{\i}a (Excellency project P07-FQM-02517) and the Royal Society (London) for financial support.

\end{document}